\documentstyle[12pt,psfig]{article}

\begin{document}

\title{The Microwave Background
Bispectrum  \\
Paper I: Basic Formalism}
\author{David N. Spergel and David M. Goldberg \\
Princeton University Observatory, Princeton, NJ\ 08544}
\maketitle

\begin{abstract}
In this paper, we discuss the potential importance of measuring the
CMB anisotropy bispectrum.  We develop a formalism for computing the
bispectrum and for measuring it from microwave background maps.  As an
example, we compute the bispectrum resulting from the 2nd order
Rees-Sciama effect, and find that is undetectable with current and
upcoming missions.
\end{abstract}

\section{Introduction}
Observations of microwave background fluctuations are a powerful probe
of the physical conditions in the early universe.  Most analyses of
the microwave background focus on measuring and interpreting the
two-point function.  If the microwave background fluctuations were a
purely Gaussian random field, then the two-point function would
completely characterize the microwave background. The non-linear
growth of structure produces non-Gaussian fluctuations.  If
detectable, these non-Gaussian fluctuations are a new window into the
evolution and growth of structure.

The three-point function and its analog, the bispectrum, are potential
tools for detecting these non-linear effects.  At present, there is no
clear signature of such modes.  There was no detection of three point
correlations in the COBE data, nor any topological signatures of
non-Gaussianity\cite{Kogut96}. While there is a reported detection of
non-Gaussianity in the COBE bispectrum\cite{Ferreira98,Heavans98}.
Our ability to detect these effects will soon improve by many orders
of magnitude with the launch of the MAP satellite\cite{MAP} and the
PLANCK\cite{PLANCK} satellite.  These upcoming launches motivate our
detailed study of the bispectrum and the three-point function.

In this article, we first discuss why the bispectrum is interesting
(\S~\ref{sec:threept}), develop a general form for calculating the
bispectrum (\S~\ref{sec:bispec}) and then discuss the particular case
of those perturbations resulting from the Rees-Sciama effect
(\S~\ref{sec:rs}).  While the Rees-Sciama bispectrum is not
detectable, we show in a companion paper that the non-Gaussian effects
produced at low redshift through a coupling of gravitational lensing
and either the Sunyaev-Zel'dovich effect or the late time ISW effect
produces a detectable signal.

\section{Why is the Bispectrum Interesting?}
\label{sec:threept}

The three-point function and its transform, the bispectrum, are
potential tools for detecting non-Gaussianity in the microwave sky.
Just as the multipole spectrum is a useful statistic for studying the
two-point correlation, the bispectrum is useful in studying the
three-point function.  To put this in context, we define the $a_{lm}$
as the expansion of the temperature anisotropies into spherical
harmonics:
\begin{equation}
a_{lm}\equiv \int d\hat{\bf n} T(\hat{\bf n}) Y_{lm}^\ast(\hat{\bf n})
\end{equation}
Thus, since the three point function is defined as $B(\hat{\bf
l},\hat{\bf m},\hat{\bf n})\equiv T(\hat{\bf l}) T(\hat{\bf m})
T(\hat{\bf n})$, we define the bispectrum as:
\begin{equation}
B_{l_1 l_2 l_3}^{m_1 m_2 m_3}=a_{l_1 m_1}a_{l_2 m_2}a_{l_3 m_3}
\end{equation}

The universe is thought to be rotationally invariant, and hence, a
more useful quantity is the angle-averaged bispectrum:
\begin{equation}
B_{l_1l_2l_3}=\sum_{m_1,m_2,m_3}\left( 
\begin{array}{ccc}
l_1 & l_2 & l_3 \\ 
m_1 & m_2 & m_3
\end{array}
\right) 
B_{l_1 l_2 l_3}^{m_1 m_2 m_3}
\label{eq:angave}
\end{equation}

Even if there was no non-Gaussianity in the initial potential
fluctuations, non-linear physics will produce mode-mode couplings.  We
argue here that the bispectrum is a particularly powerful tool for
detecting these couplings.

In inflation, quantum fluctuations produce Gaussian random phase variations
in the gravitational potential: 
\begin{equation}
\Phi ({\bf x})=\int \frac{d^3{\bf k}}{\left( 2\pi \right) ^3}
\Phi_0({\bf k})\exp(i{\bf k\cdot x})
\end{equation}
with $\langle \Phi_0({\bf k})\Phi_0^{*}({\bf k}^{\prime })\rangle
=P(k)\delta ^{(3)}( {\bf k-k}')$, the power spectrum of potential
fluctuations at the current epoch. While processes during inflation
may produce some non-Gaussianity, this is small in most inflationary
models \cite{Falk93}. On the other hand, when the fluctuations reenter
the horizon, there are many non-linear effects that alter these
initial conditions.  Near decoupling, non-linear hydrodynamical
effects couple modes \cite{Peebles80}. After decoupling, non-linear
general relativistic corrections also alter the microwave background
spectrum \cite{Pyne96, Mollerach97} as do non-linear coupling between
density fluctuations\cite{Mollerach95,Munshi95}. However, since
potential fluctuations in the early universe are small, $\langle
\Phi^2 \rangle \sim 10^{-9},$ these non-linear effects are usually
viewed as undetectable.  For example, Mollerach et
al.\cite{Mollerach95} and Munshi et al.\cite{Munshi95} estimated the
reduced bispectrum generated by non-linear evolution of long
wavelength modes and showed that they are undetectable in the COBE
maps.  In a spatially flat universe, the microwave background
temperature at the surface of last scatter can be linearly related to
the potential fluctuations at that epoch, giving us what we shall
henceforth call the linear temperature approximation:
\begin{equation}
T^{L}(\hat{\bf n})=\phi(\tau_{r})
\int \frac{d^{3}{\bf k}}{(2\pi)^{3}}e^{i{\bf k}\cdot\hat{\bf n}\tau_r}
\Phi_{0}({\bf k}) g(k),
\end{equation}
where $\tau$ is the comoving (conformal) lookback time, $\phi(\tau)$
is a normalization constant for potential perturbations which is equal
to unity today, and $g(k)$ is the linear radiation transfer function
and describes the relationship between potential fluctuations and
temperature fluctuations at recombination.  For small $k$, $g(k) =
1/3$.  For larger $k$, we need to evolve the coupled matter-radiation
fluid.  Codes like CMBFAST compute $g(k)$ for different cosmological
models.

Expanding the linear part of the temperature perturbation into
spherical harmonics:
\begin{equation}
a^L_{lm}=\phi(\tau_r)\int \frac{d^3{\bf k}}{(2\pi)^{3}}
d\hat{\bf n} e^{i{\bf k}\cdot\hat{\bf n}\tau}\Phi_0({\bf k}) g(k)
Y_{lm}^\ast(\hat{\bf n})\ .
\end{equation}

Since we will later wish to integrate over the sky, it will be useful to expand
out the exponential into spherical harmonics as well using the Rayleigh
expansion:
\begin{equation}
e^{i{\bf k}\cdot{\bf r}}=4\pi\sum_{lm}i^lj_l(kr)Y_{lm}^\ast(\hat{\bf r})
Y_{lm}(\hat{\bf k})\ .
\end{equation}

Thus, the linear approximation of the CMB fluctuations are given by:
\begin{equation}
a^L_{lm}=4\pi \phi(\tau_r) (-i)^l \int \frac{d^3{\bf k}}{(2\pi)^{3}}
\Phi_0(\vec{k}) Y_{l,-m}(\hat{\bf k})j_l(k\tau_r)g(k)
\label{eq:alin}
\end{equation}

In addition to the terms proportional to $\Phi_0({\bf k})$ (linear),
there are a number of terms in the temperature fluctuations which are
proportional to $\Phi_0({\bf k})^2$ (nonlinear).  These corrections
arise from gravitational couplings, gravitational lensing, and
radiation effects.  These terms generically add a correction to the
microwave background:
\begin{eqnarray}
T^{NL}(\hat{\bf n})&=& \int_{\tau_r}^{0}d\tau \Phi^{2}(\hat{\bf n})
f(\hat{\bf n}\tau,\tau) \\ \nonumber
&=&\int d\tau \phi^2(\tau)
\int \frac{d^3{\bf k}_1}{(2\pi)^3}
\frac{d^3{\bf k}_2}{(2\pi)^3}
\Phi_0({\bf k}_1)\Phi_0({\bf k}_2)
e^{i({\bf k}_1+{\bf k}_2)\cdot \hat{\bf n}\tau}f({\bf k}_1,{\bf k}_2,\tau)
\end{eqnarray}

We may again take the spherical harmonic transform of the non-linear
temperature perturbation to calculate the coefficient, $a_{lm}^{NL}$.
We will find it convenient to expand the coupling term, $f({\bf
k}_1,{\bf k}_2,\tau)$ into Legendre polynomials:
\begin{equation}
f({\bf k}_1,{\bf k}_2,\tau)\equiv
\sum_{l}f_{l}(k_1,k_2,\tau)P_{l}(\hat{\bf k}_1\cdot\hat{\bf k}_2)=
\sum_{l m}\frac{4\pi}{2l+1}f_l(k_1,k_2,\tau)Y_{lm}(\hat{\bf k}_1)
Y_{lm}^\ast(\hat{\bf k}_2)
\end{equation}

Since the non-linear term will only couple to itself if ${\bf k}_1={\bf k}_2$,
we can safely ignore this effect, and instead concentrate on the case in which
a bispectrum is produced by:
\begin{equation}
B^{l_1 l_2 l_3}_{m_1 m_2 m_3}=
a^{L}_{l_1 m_1}a^{L}_{l_2 m_2}a^{NL\ast}_{l_3 m_3}+
a^{L}_{l_2 m_2}a^{L}_{l_3 m_3}a^{NL\ast}_{l_1 m_1}+
a^{L}_{l_3 m_3}a^{L}_{l_1 m_1}a^{NL\ast}_{l_2 m_2}
\label{eq:term1}
\end{equation}
We will only examine the first term in this derivation, and add in the other
two permutation in the end.

Looking at $a^{NL\ast}_{l_3 m_3}$, we find:
\begin{eqnarray}
a_{l_3 m_3}^{NL\ast} &=&
\frac{(4\pi)^3}{(2\pi)^6}
\int d\tau \phi^2(\tau)
\int k_1^2 dk_1 k_2^2 dk_2
\Phi_0^\ast({\bf k}_1)\Phi_0^\ast({\bf k}_2) \\ \nonumber
&\times&
\sum_{l l' l'' m m' m''} (-i)^{l'+l''}\frac{1}{2l+1}
f_l(k_1,k_2,\tau) j_{l'}(k_1\tau) j_{l''}(k_2\tau) \\ \nonumber
&\times&
\int d\hat{\bf n} d\hat{\bf k}_1 d\hat{\bf k}_2
Y_{l_3 m_3}(\hat{\bf n}) Y_{l'm'}(\hat{\bf n}) Y_{l'' m''}(\hat{\bf n})
Y_{l'm'}^{\ast}(\hat{\bf k}_1) Y_{lm}^{\ast}(\hat{\bf k}_1)
Y_{lm}(\hat{\bf k}_2) Y_{l''m''}^\ast(\hat{\bf k}_2)
\end{eqnarray}

Substituting this expression and equation~(\ref{eq:alin}) into
equation~(\ref{eq:term1}), we get:
\begin{eqnarray}
a^{L}_{l_1 m_1}a^{L}_{l_2 m_2}a^{NL\ast}_{l_3 m_3}&=&
\frac{(4\pi)^5}{(2\pi)^6}\phi^2(\tau_r) \int d\tau \phi^2(\tau)
\int \frac{d^3{\bf k}}{(2\pi)^3} \frac{d^3{\bf k}'}{(2\pi)^3} 
\int k_1^2 dk_1 k_2^2 dk_2 \\ \nonumber
&\times&
\Phi_0({\bf k})\Phi_0({\bf k}')\Phi_0^\ast({\bf k}_1)\Phi_0^\ast({\bf k}_2)
g(k) g(k')\\ \nonumber
&\times&
\sum_{l l' l'' m m' m''} \frac{1}{2l+1}
f_l(k_1,k_2,\tau) j_{l'}(k_1\tau) j_{l''}(k_2\tau)
j_{l_1}(k\tau_r)j_{l_2}(k'\tau_r)
\\ \nonumber
&\times&
\int d\hat{\bf n} d\hat{\bf k}_1 
Y_{l_3 m_3}(\hat{\bf n}) Y_{l'm'}(\hat{\bf n}) Y_{l'' m''}(\hat{\bf n})
Y_{l'm'}^{\ast}(\hat{\bf k}_1) Y_{lm}^{\ast}(\hat{\bf k}_1)
Y^\ast_{l_1 m_1}(\hat {\bf k})\\ \nonumber
&\times&
\int d\hat{\bf k}_2
Y_{lm}(\hat{\bf k}_2) Y_{l''m''}^\ast(\hat{\bf k}_2)
Y_{l_2 m_2}^\ast(\hat{\bf k}')
\end{eqnarray}

The integrals over ${\bf k}$ and ${\bf k}'$ give two possible
permutations, $\delta({\bf k}-{\bf k}_1)$ and $\delta({\bf k}-{\bf
k}_2)$.  This is patently symmetric, and thus, we will chose the
former and multiply the integral by 2.  After integrating over ${\bf
k}$ and ${\bf k}'$, we find that we have 3 sets of 3 spherical
harmonics.  This is precisely the form of the $Q$ coefficients
discussed in the appendix.  Thus, integrating over $\hat{\bf n}$,
$\hat{\bf k}_1$, and $\hat{\bf k}_2$, summing over $m,m',m''$ , and
convolving with the Wigner 3-j symbol over $m_1,m_2,m_3$ as in
equation~(\ref{eq:angave}), we find that the first of three terms in the
angle averaged bispectrum is:
\begin{eqnarray}
B^{(1)}_{l_1 l_2 l_3}&=&\frac{1}{2\pi^4}\phi^2(\tau_r)
\int d\tau \phi^2(\tau)\int dk_1 dk_2 k_1^2 k_2^2 P(k_1)P(k_2)
g(k_1)g(k_2)
\\ \nonumber
&\times&
\sum_{l l' l''}(2l'+1)(2l''+1)Q_{ll''l'}^{l_1 l_2 l_3}
f_l(k_1,k_2,\tau)j_{l'}(k_1\tau)j_{l''}(k_2\tau)
j_{l_1}(k_1\tau_r)j_{l_2}(k_2\tau_r)
\end{eqnarray}

\section{Measuring the Bispectrum}
\label{sec:bispec}

Measuring this effect from observations is quite straightforward.  If
there is complete sky coverage and no sky cuts, the the angle averaged
bispectrum can be computed by rewriting the Wigner 3-j symbols as
integrals over the sky,

\begin{eqnarray}
\left( 
\begin{array}{ccc}
l_1 & l_2 & l_3 \\ 
0 & 0 & 0
\end{array}
\right) B_{l_1l_2l_3} &=&\sum_{m_1,m_2,m_3}^{}\left( 
\begin{array}{ccc}
l_1 & l_2 & l_3 \\ 
0 & 0 & 0
\end{array}
\right) \left( 
\begin{array}{ccc}
l_1 & l_2 & l_3 \\ 
m_1 & m_2 & m_3
\end{array}
\right) a_{l_1m_1}a_{l_2m_2}a_{l_3m_3}  \nonumber \\
&&  \nonumber \\
\  &=&\sqrt{\frac{\left( 4\pi \right) ^3}{(2l_1+1)(2l_2+1)(2l_3+1)}} 
\nonumber \\
&&\ \qquad \times \sum_{m_1,m_2,m_3}^{}\int d{\bf \hat q}Y_{l_1m_1}({\bf %
\hat q})Y_{l_2m_2}^{}({\bf \hat q)}Y_{l_3m_3}({\bf \hat q)}  \nonumber \\
&&\ \qquad \times \int d{\bf \hat l}d{\bf \hat m}d{\bf \hat n}Y_{l_1m_1}^{*}(%
{\bf \hat l})Y_{l_2m_2}^{*}({\bf \hat m)}Y_{l_3m_3}^{*}({\bf \hat n)}T({\bf %
\hat l)}T({\bf \hat m)}T({\bf \hat n)}  \nonumber \\
\  &=&\int d{\bf \hat q}e_{l_1}({\bf \hat q})e_{l_2}({\bf \hat q})e_{l_3}(%
{\bf \hat q})
\end{eqnarray}
where the sky map is averaged over rings centered around point ${\bf \hat q}$%
: 
\begin{equation}
e_l({\bf \hat q})=\sqrt{\frac{2l+1}{4\pi }}\int d{\bf \hat l}T({\bf \hat l)}%
P_l\left( {\bf \hat q\cdot \hat l}\right) 
\end{equation}

Since the bispectrum signal is rather weak, the dominant source
of noise is the cosmic variance of the dominant Gaussian signal.
For a Gaussian random field, $\langle a_{l_1m_1}a_{l_2m_2}a_{l_3m_3}\rangle
=0.$ However, its variance, $\langle
(a_{l_1m_1}a_{l_2m_2}a_{l_3m_3})^2\rangle =c_{l_1}c_{l_2}c_{l_3}$ for $%
l_1\ne l_2$, $l_2\ne l_3$, $l_1\ne l_3$\cite{Luo94}. When we include
detector noise, then 
\begin{equation}
\langle (a_{l_1m_1}a_{l_2m_2}a_{l_3m_3})^2\rangle =\left( c_{l_1}+\sigma
_0^2w_{l_1}^{-2}\right) \left( c_{l_2}+\sigma _0^2w_{l_2}^{-2}\right) \left(
c_{l_3}+\sigma _0^2w_{l_3}^{-2}\right) 
\end{equation}
where $\sigma _0$ is the detector noise and $w_l$ is the experimental window
function. Similarly, for a Gaussian field, 
\begin{eqnarray}
\left\langle B_{l_1l_2l_3}\right\rangle  &=&0  \nonumber   \\
\left\langle B_{l_1l_2l_3}B_{l_1^{\prime }l_2^{\prime }l_3^{\prime
}}\right\rangle  &=&\left( c_{l_1}+\sigma _0^2w_{l_1}^{-2}\right) \left(
c_{l_2}+\sigma _0^2w_{l_2}^{-2}\right) \left( c_{l_3}+\sigma
_0^2w_{l_3}^{-2}\right)   \label{B2} \\
&&\qquad \times \left( \delta _{l_1l_2l_3}^{l_1^{\prime }l_2^{\prime
}l_3^{\prime }}+\delta _{_{l_1l_3l_2}}^{l_1^{\prime }l_2^{\prime
}l_3^{\prime }}+\delta _{l_2l_1l_3}^{l_1^{\prime }l_2^{\prime }l_3^{\prime
}}+\delta _{l_2l_3l_1}^{l_1^{\prime }l_2^{\prime }l_3^{\prime }}+\delta
_{l_3l_1l_2}^{l_1^{\prime }l_2^{\prime }l_3^{\prime }}+\delta
_{l_3l_2l_1}^{l_1^{\prime }l_2^{\prime }l_3^{\prime }}\right) \nonumber  
\end{eqnarray}
for $l_1\ne l_2$, $l_2\ne l_3$, $l_1\ne l_3.$ Because of this symmetry, we
restrict ourselves to bispectrum terms where $l_1<l_2<l_3.$ 

\section{The Bispectrum from the Rees-Sciama Effect}
\label{sec:rs}

Several groups\cite{Mollerach95,Munshi95} have discussed
the possibility of detecting the non-linear signature
of the Rees-Sciama effect in the COBE data.  They concluded
that the signal was undetectable on COBE scales. In their
analysis, they focus only on the diagonal terms in the bispectrum Here,
we compute the predicted signal on smaller angular scales and
compute all of the bispectrum terms.

The second order Rees-Sciama effect arises from
the non-linear growth of density fluctuations\cite{Mollerach95}:
\begin{eqnarray}
T^{NL}(\hat{\bf n})&=&2\int d\tau \frac{\partial}{\partial \tau}
\Phi^{NL}(\hat{\bf n}\tau,\tau)\\ \nonumber
&=&2\int d\tau \frac{\partial}{\partial \tau}
\int \frac{d^3{\bf k}}{(2\pi)^3}\Phi^{NL}(\hat{\bf k},\tau)
e^{i{\bf k}\cdot \hat{\bf n}\tau}
\end{eqnarray}
and give the second-order potential perturbation as:
\begin{equation}
\Phi^{NL}({\bf k},\tau)=-\frac{(\tau_0-\tau)^2}{84k^2}
\phi^2(\tau)
\int \frac{(d^3{\bf k}')}{(2\pi)^{3}}
\Phi_0({\bf k}-{\bf k}')\Phi_0({\bf k'})
\left[
3k^2k'^2+7k^2{\bf k}\cdot{\bf k}'-10({\bf k}\cdot{\bf k}')^2
\right]
\end{equation}
This second -order approximation is valid for $k < 1h$ Mpc$^{-1}$.
\cite{Seljak96b}.

Since we expect much of the signal to come from high redshift
contributions, and since in a flat, $\Omega_m=1$ universe,
$\dot\phi(\tau)=0$, we will take $\phi(\tau)$ is constant for the
remainder of this derivation.  Additionally, we will redefine our
variables of integration: ${\bf k}_1\equiv{\bf k}-{\bf k}'$ and ${\bf
k}_2={\bf k}'$.  This yields:
\begin{eqnarray}
T^{NL}(\hat{\bf n})&=&\int d\tau
\phi^2(\tau)
\int \frac{d^3{\bf k}_1}{(2\pi)^3}\frac{d^3{\bf k}_2}{(2\pi)^3}
\Phi_0({\bf k}_1)\Phi_0({\bf k}_2)e^{i({\bf k}_1+{\bf k}_2)\hat{\bf n}\tau}
\\ \nonumber
&\times&
\frac{(\tau_0-\tau)}{21({\bf k}_1+{\bf k}_2)^2}
\left[
10k_1^2k_2^2+7(k_1^2+k_2^2)k_1k_2\mu+4k_1^2k_2^2\mu^2
\right]
\end{eqnarray}
where $\mu\equiv \hat{\bf k}_1\cdot\hat{\bf k}_2$.

The second part of the equation is the coupling function, $f(\hat{\bf
k}_1,\hat{\bf k}_2, \tau)$. The vectors in the denominator make it
difficult to decompose it into Legendre polynomials, however, this may
be simplified by noting that:
\begin{equation}
a_{lm}^{NL}\propto\int d\hat{\bf n}e^{i({\bf k}_1+{\bf k}_2)\hat{\bf n}\tau}
Y^\ast_{lm}(\hat{\bf n})
\end{equation}
Thus, taking the Laplacian of $a_{lm}^{NL}$, we find:
\begin{eqnarray}
l(l+1)a_{lm}^{NL\ast}
&=&\int d\hat{\bf n} \int d\tau \phi^2(\tau)
\int \frac{d^3{\bf k}_1}{(2\pi)^3}
\frac{d^3{\bf k}_2}{(2\pi)^3}
\Phi_0({\bf k}_1)\Phi_0({\bf k}_2)
e^{i({\bf k}_1+{\bf k}_2)\cdot \hat{\bf n}\tau}Y^{NL}_{lm}(\hat{\bf n})
\nonumber \\
&\times& f({\bf k}_1,{\bf k}_2,\tau)
({\bf k}_1+{\bf k}_2)^2\tau^2
\end{eqnarray}

We now define a new coupling function:
\begin{equation}
\tilde{f}({\bf k}_1,{\bf k}_2,\tau)\equiv f({\bf k}_1,{\bf k}_2,\tau)
({\bf k}_1+{\bf k}_2)^2\tau^2\ ,
\end{equation}
which can be easily substituted into the definition of the bispectrum, with
$B_{l_1 l_2 l_3}\Rightarrow l_3 (l_3+1)B_{l_1 l_2 l_3}$ and $f_{l}\Rightarrow
\tilde{f}_l$.

For the Rees-Sciama effect, this is decomposed into only the $l=0$,$l=1$, and
$l=2$ Legendre coefficients:
\begin{eqnarray}
\tilde{f}_{0}(k_1,k_2,\tau)&=&\frac{2 (\tau_0-\tau)\tau^2}{3}k_1k_2
\left(\frac{17}{21}k_1 k_2\right) \\
\tilde{f}_{0}(k_1,k_2,\tau)&=&\frac{2 (\tau_0-\tau)\tau^2}{3}k_1k_2
\left(\frac{1}{2}(k_1^2+k_2^2)\right) \\
\tilde{f}_{2}(k_1,k_2,\tau)&=&\frac{2 (\tau_0-\tau)\tau^2}{3}k_1k_2
\left(\frac{4}{21}k_1 k_2\right) \\
\end{eqnarray}

so,
\begin{eqnarray}
l_3 (l_3+1) B_{l_1 l_2 l_3}&=&\frac{1}{3\pi^4}\phi^2(\tau_r)
\int d\tau \phi^2(\tau)(\tau_0-\tau)\tau^2 \\ \nonumber
&\times& \int dk_1 dk_2 k_1^3 k_2^3 P(k_1) P(k_2) g(k_1) g(k_2)
j_{l_1}(k_1 \tau_r) j_{l_2}(k_2 \tau_r) \\ \nonumber
&\times& \sum_{l'l''}j_{l'}(k_1 \tau)j_{l''}(k_2 \tau)
(2l'+1)(2l''+1) \\ \nonumber
&\times& \left[
\frac{17}{21} Q_{0 l'' l'}^{l_1 l_2 l_3} k_1 k_2 +
\frac{1}{2} Q_{1 l'' l'}^{l_1 l_2 l_3} (k_1^2+k_2^2) +
\frac{4}{21} Q_{2 l'' l'}^{l_1 l_2 l_3} k_1 k_2
\right]
\end{eqnarray}

As discussed in the appendix, $\sum_{l' l''}(2l'+1)(2l''+1)Q_{ll''l'}^{l_1 l_2
l_3}=(2l_1+1)(2l_2+1)Q_{0 l_1 l_2}^{l_1 l_2 l_3}$, and thus, if $l$ is small (in this case, it
only takes on the values 0,1,2), and $h_{l'l''}$ is a function which varies
slowly with $l'$ and $l''$, we may make the approximation:
\begin{equation}
\sum_{l' l''}(2l'+1)(2l''+1)Q_{ll''l'}^{l_1 l_2 l_3}h_{l'l''}
\simeq (2l_1+1)(2l_2+1) Q_{0 l_1 l_2}^{l_1 l_2 l_3}h_{l_2 l_1}
\end{equation}
Under this assumption, the bispectrum reduces to:
\begin{eqnarray}
l_3 (l_3+1) B_{l_1 l_2 l_3}
&=&\frac{4\phi^2(\tau_r)I_{l_1 l_2 l_3}}{3\pi^{5/2}}
\int d\tau \phi^2(\tau)(\tau_0-\tau)\tau^2 \\ \nonumber
&\times& \int dk_1 dk_2 k_1^3 k_2^3 P(k_1) P(k_2) g(k_1) g(k_2)
\\ \nonumber
&\times&  j_{l_1}(k_1 \tau_r) j_{l_2}(k_2 \tau_r) 
j_{l_2}(k_1 \tau)j_{l_1}(k_2 \tau) 
(k_1+k_2)^2
\end{eqnarray}
where
\begin{equation}
I_{l_1 l_2 l_3} = \sqrt{(2l_1+1)(2l_2+1)(2l_3+1)}
\left( 
\begin{array}{ccc}
l_1 & l_2 & l_3 \\ 
0 & 0 & 0
\end{array}
\right)
\end{equation}

In order to deal with the spherical Bessel functions, we will first note that
in the asymptotic limit, if $x > l$, $j_{l}(x)\simeq \sin(x-\pi l/2)/x$.  The
lower limit of integration will be taken to be:
\begin{equation}
\tau_{min}={\rm max}\left( \frac{l_1}{k_2},\frac{l_2}{k_1}\right)
\end{equation}
Likewise, the lower limit of integration on $k_1$ is $l_1/\tau_r$ and $k_2$ is
$l_2/\tau_r$.  In this limit, we can expand out the four spherical Bessel
functions as:
\begin{equation}
j_{l_1}(k_1 \tau_r) j_{l_2}(k_2 \tau_r)
j_{l_2}(k_1 \tau)j_{l_1}(k_2 \tau) \simeq
\frac{1}{4}(-1)^{l_1+l_2}
\frac{\cos\left[(k_1-k_2)\tau_r)\right]\cos\left[(k_1-k_2)\tau)\right]
}{k_1^2k_2^2\tau^2 \tau_r^2}
\end{equation}
plus additional terms which oscillate much more quickly.  

We can now approximate the the time integral:
\begin{eqnarray}
\int_{\tau_{min}}^{\tau_r} d\tau \tau^2  (\tau_0-\tau) \phi^2(\tau) 
j_{l_1}(k_1 \tau_r) j_{l_2}(k_2 \tau_r)
j_{l_2}(k_1 \tau)j_{l_1}(k_2 \tau)\\ \nonumber
\simeq \frac{1}{4}(-1)^{l_1+l_2}\cos[(k_1-k_2)\tau_r]
\int_{\tau_{min}}^{\tau_r} d\tau (\tau_0-\tau) \phi^2(\tau)
\frac{\cos\left[(k_1-k_2)\tau\right]}{k_1^2k_2^2\tau_r^2} \\ \nonumber
\simeq \frac{\pi}{4}\frac{\delta(k_1-k_2)}{k_1^2 k_2^2\tau_r^2}
(-1)^{l_1+l_2}
\phi^2(\tau_{min})(\tau_0-\tau_{min})
\end{eqnarray}
where we have approximated $\phi(\tau)(\tau_0-\tau)$ as a slowly varying
function which is maximally weighted near $\tau=\tau_{min}$, and $\sin(x)/x$ is
approximated as a delta function.

Thus, the bispectrum can be approximated as:
\begin{equation}
B^{(1)}_{l_1 l_2 l_3}\simeq\frac{\phi^{2}(\tau_r)}{\tau_r^2}
\frac{4}{3 \pi^{3/2}}
\frac{I_{l_1 l_2 l_3}}{l_3(l_3+1)}
\int^{\infty}_{l_m/\tau_r} dk k^4 P^2(k) g^2(k)
\phi^2(\tau_{min})(\tau_0-\tau_{min})
\end{equation}
where we have defined $l_m\equiv max(l_1,l_2)$.  

In equation~(\ref{eq:alin}), we defined $g(k)$ in terms of
$a_{lm}^L$.  Defining $c_l\equiv \langle a_{lm}^L
a_{lm}^{L\ast}\rangle$, we
find:
\begin{equation}
c_l=\frac{2}{\pi} \int k^2 dk P(k) j_l^2(k\tau_r) g^2(k)
\end{equation}

\begin{figure}[tbp]
\centerline{\psfig{figure=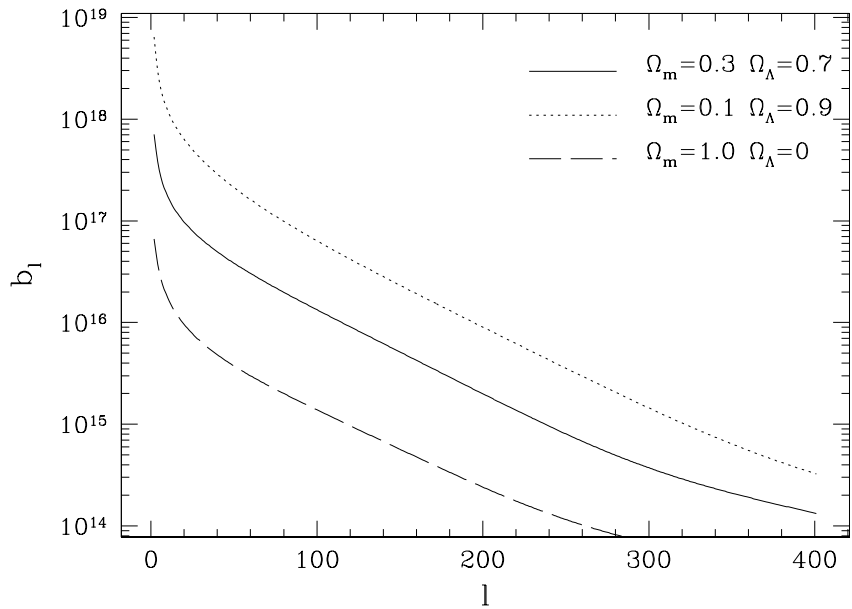,height=5.0in,angle=0}}
\caption{The coefficients $b_{l_m}$ for various cosmologies, as
defined in the text.  Note that for low $\Omega$ models, the signal
is significantly larger than $\Omega_m=1$ models at all l.}
\label{fg:bl1}
\end{figure}

For large values the spherical Bessel functions are very close to zero
for $k\tau_r < l$ and since $k^2 P(k)$ is a monotonically decreasing
function for all values of $k$, we may approximate
$j_{l}^2(x)=\pi\delta(l-x)/2$.  With this approximation:
\begin{equation}
c_{k\tau_r}\simeq
\frac{\phi^2(\tau_r)}{\tau_r(2k/\tau_r+1)}P(k)g^2(k)k^2
\end{equation}
Substituting this into the previous expression yields:
\begin{eqnarray}
B^{(1)}_{l_1 l_2 l_3}&\simeq&\frac{4}{3\pi^{3/2}\tau_r}
\frac{I_{l_1 l_2 l_3}}{l_3(l_3+1)}
\int_{k_{min}}^{\infty}dk k^2 P(k)c_{k\tau_r}(2k\tau_r+1)
\phi^2(\tau_{min})(\tau_0-\tau_{min})\nonumber \\
&\simeq&
\frac{4}{3\pi^{3/2}\tau_r^4}
\frac{I_{l_1 l_2 l_3}}{l_3(l_3+1)}
\sum_{l=l_m} l^2 (2l+1) P\left(\frac{l}{\tau_{r}}\right)c_l
\phi^2(\tau_r l_m/l)
(\tau_0-\tau_r l_m/l) \nonumber \\ 
&\equiv&
\frac{4}{3\pi^{3/2}\tau_r^4}
\frac{I_{l_1 l_2 l_3}}{l_3(l_3+1)}
b_{l_m}
\end{eqnarray}
In Figure~\ref{fg:bl1} we have plotted the coefficient $b_{l_m}$ for
flat cosmologies where $\Omega_{m}=0.1$, $0.3$, and $1.0$.  

Will we be able to detect this effect? We can estimate this by determining
whether in a $\Omega _m=0.3,\Omega _\Lambda =0.7$ universe, we can use the
MAP data to reject the hypothesis that there is no non-Gaussianity in the
CMB maps: 
\begin{equation}
\chi ^2=\sum_{l_1,l_2,l_3}{\frac{\langle B_{l_1,l_2,l_3}\rangle ^2}{\langle
B_{l_1,l_2,l_3}^2\rangle }}
\end{equation}

\begin{figure}[tbp]
\centerline{\psfig{figure=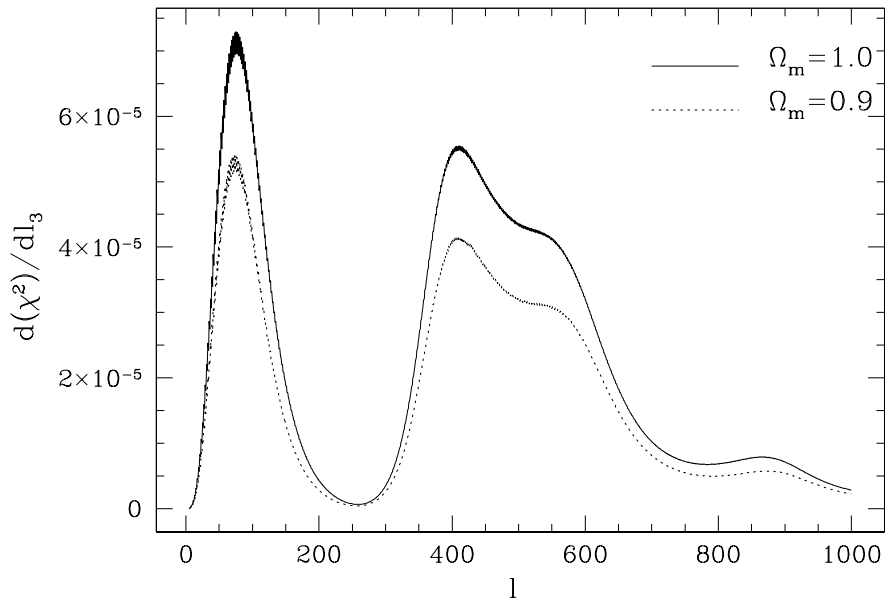,height=5.0in,angle=0}}
\caption{{}The amount of information gained by increasing $l_3$, as
given by $\chi^2=\sum (\langle B_{l_1 l_2 l_3}(\Omega_m)\rangle
-\langle B_{l_1 l_2 l_3}(\Omega_m=0.3)\rangle )^2/\langle B_{l_1 l_2
l_3}^2\rangle$, where we have used the PLANCK detection sensitivity.
Since there is very little contribution above $l_3=700$, the MAP and
PLANCK missions will be equally sensitive to this effect.  Our
fiducial model here and throughout is $\Omega_m=0.3$,
$\Omega_{\Lambda}=0.7$.  The shape of this spectrum comes from several
competing effects.  At large $l$, more modes are included in the sum,
providing a larger signal.  However, the signal per mode becomes
increasingly weak at a logarithmic rate for high l, as suggested by
the plot of the coefficients in Figure 1.  Finally, the features in
the spectrum above are caused by variations in the expected noise.
Troughs correspond to peaks in the CMB spectrum.  }
\label{fg:dchi}
\end{figure}

In Figure~\ref{fg:dchi} we show the amount of ``information'' gained
by increasing the maximum index ($l_3$ in our notation).  Note that
since all of the signal occurs around $l<600$, and the MAP and PLANCK
satellites are cosmic variance dominated up to $l\simeq 600$ and
$l\simeq 1000$ respectively, both experiments should measure the
effect with approximately the same sensitivity.

\begin{figure}[tbp]
\centerline{\psfig{figure=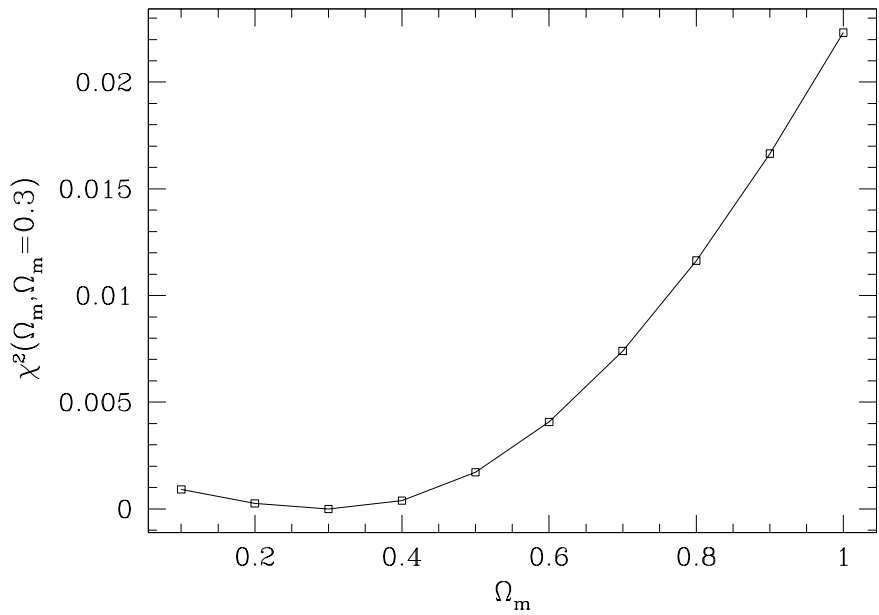,height=5.0in,angle=0}}
\caption{{}The total value of $\chi^2(\Omega_m,\Omega_m=0.3)$ for
variations in $\Omega_m$ for both PLANCK and MAP.  All values of
$\Omega_m$ produce signals which are consistent with $\Omega_m=0.3$.}
\label{fg:chitot}
\end{figure}

The integral of Figure~\ref{fg:dchi} gives us the total $\chi^2$
between our fiducial model and some test model.  In
Figure~\ref{fg:chitot}, we plot the total $\chi^2$ under the
assumption that the ``true'' universe is a flat $\Omega_m=0.3$, and
the test models remain flat and vary $\Omega_m$.  Even in the extreme
case in which we assume $\Omega_m=0.3$ and the true universe has
$\Omega_m=1.0$, the $\chi^2$ for PLANCK is only $\sim 0.02$.

\section{Discussion}
Most previous discussions of the microwave background have focused on
multipole spectrum of the microwave background: 
\begin{equation}
c_l = {\frac{1}{2l+1}} \sum_{m} a_{lm} a^*_{lm}
\end{equation}
Because of azimuthal symmetry, the expectation value of $\langle a_{lm}^2
\rangle$ is independent of $m$. If the microwave background fluctuations
were purely a Gaussian random field, then the multipole spectrum measures
all of the statistical properties of the microwave background.

In this paper, we have developed formalism for computing the
bispectrum and for measuring it from microwave background maps.  As an
example, we computed the Rees-Sciama bispectrum and found that is
undetectable with current and forthcoming missions.  In a companion
paper\cite{Goldberg98}, we will work out the bispectrum for nonlinear
contributions from the coupling of gravitational lensing and
low-redshift temperature perturbations.  There, we show that MAP will
be able to detect the SZ-lensing bispectrum and PLANCK will be able to
detect the ISW-lensing bispectrum.

Much work remains to be done on the bispectrum. We do not have a
robust statistical technique for computing the bispectrum from a
million pixel map with sky cuts and spatially variable noise, nor do
we understand how foregrounds will contaminate measurements of the
bispectrum.  For example, with MAP, radio sources are expected to
produce marginally detectable skewness\cite{fridge}.  There are also
other physical effects that may have observable signatures in the
bispectrum.

\section{Acknowledgments}

We thank Martin Bucher, Jeremy Goodman, Gary Hinshaw, Arthur Kosowsky and Jim
Peebles for useful discussions. DNS acknowledges the MAP/MIDEX project for
support. DMG is supported by an NSF graduate research fellowship.

\appendix
\section{Wigner Symbol Identities}

In this appendix, we derive some relevant properties for the integral,

\begin{eqnarray}
Q_{ll^{\prime }l^{\prime \prime }}^{l_1l_2l_3} &\equiv& \int d\hat{{\bf l}} d%
\hat{{\bf m}} d\hat{{\bf n}} P_l(\cos \alpha )P_{l^{\prime }}(\cos \beta
)P_{l^{\prime \prime }}(\cos \gamma )  \nonumber \\
&&\qquad \qquad \qquad \qquad \times \sum_{m_1,m_2m_3}\left( 
\begin{array}{ccc}
l_1 & l_2 & l_3 \\ 
m_1 & m_2 & m_3
\end{array}
\right) Y_{l_1m_1}^{*}(\hat l)Y_{l_2m_2}^{*}(\hat m)Y_{l_3m_3}^{*}(\hat n) 
\nonumber \\
&=&\left( \frac{4\pi }{2l+1}\right) \left( \frac{4\pi }{2l^{\prime }+1}%
\right) \left( \frac{4\pi }{2l^{\prime \prime }+1}\right)
\sum_{m_1,m_2m_3}\left( 
\begin{array}{ccc}
l_1 & l_2 & l_3 \\ 
m_1 & m_2 & m_3
\end{array}
\right)  \nonumber \\
&&\qquad \times \sum_{m,m^{\prime },m^{\prime \prime }} \int d\hat{{\bf l}} d%
\hat{{\bf m}} d\hat{{\bf n}} Y_{lm}(\hat l) Y_{lm}^{*}(\hat m)
Y_{l^{\prime}m^{\prime}}(\hat m) Y_{l^{\prime}m^{\prime}}^{*}(\hat n)
Y_{l^{\prime \prime }m^{\prime \prime }}(\hat n)Y_{l^{\prime \prime
}m^{\prime \prime }}^{*}(\hat l)  \nonumber \\
&&\qquad \qquad \qquad \qquad \qquad \qquad \times Y_{l_1m_1}^{*}(\hat
l)Y_{l_2m_2}^{*}(\hat m)Y_{l_3m_3}^{*}(\hat n)  \nonumber \\
&=&I_{ll^{\prime }l^{\prime \prime }}^{l_1l_2l_3}\sum_{m_1,m_2m_3}\left( 
\begin{array}{ccc}
l_1 & l_2 & l_3 \\ 
m_1 & m_2 & m_3
\end{array}
\right) \sum_{m,m^{\prime },m^{\prime \prime }}\left( 
\begin{array}{ccc}
l & l^{\prime \prime } & l_1 \\ 
m & -m^{\prime \prime } & -m_1
\end{array}
\right)  \nonumber \\
&&\left( 
\begin{array}{ccc}
l^{\prime } & l & l_2 \\ 
m^{\prime } & -m & -m_2
\end{array}
\right) \left( 
\begin{array}{ccc}
l^{\prime\prime} & l^{\prime} & l_3 \\ 
m^{\prime\prime} & -m^{\prime} & -m_3
\end{array}
\right) (-1)^{(m+m^{\prime}+m^{\prime\prime})}
\end{eqnarray}
where $\cos \alpha =\hat l\cdot \hat m$, $\cos \beta =\hat m\cdot \hat n$
and $\cos \gamma =\hat l\cdot \hat n$ and 
\begin{equation}
I_{ll^{\prime }l^{\prime \prime }}^{l_1l_2l_3}=\sqrt{(4\pi
)^3(2l_1+1)(2l_2+1)(2l_3+1)}\left( 
\begin{array}{ccc}
l & l^{\prime \prime } & l_1 \\ 
0 & 0 & 0
\end{array}
\right) \left( 
\begin{array}{ccc}
l^{\prime } & l & l_2 \\ 
0 & 0 & 0
\end{array}
\right) \left( 
\begin{array}{ccc}
l^{\prime \prime } & l^{\prime} & l_3 \\ 
0 & 0 & 0
\end{array}
\right)
\end{equation}
We can use the definition of the Wigner 6-j symbol\cite{Varshalovich} to
rewrite the integral as, 
\begin{eqnarray}
Q_{ll^{\prime }l^{\prime \prime }}^{l_1l_2l_3} &=&I_{ll^{\prime }l^{\prime
\prime }}^{l_1l_2l_3}\left\{ 
\begin{array}{ccc}
l_1 & l_2 & l_3 \\ 
l^{\prime} & l^{\prime\prime} & l
\end{array}
\right\} \left( -1\right) ^{l+l^{\prime }+l^{\prime \prime
}}\sum_{m_1,m_2m_3}\left( 
\begin{array}{ccc}
l_1 & l_2 & l_3 \\ 
m_1 & m_2 & m_3
\end{array}
\right) ^2\left( -1\right) ^{l_1+l_2+l_3}  \nonumber \\
&=&I_{ll^{\prime }l^{\prime \prime }}^{l_1l_2l_3}\left\{ 
\begin{array}{ccc}
l_1 & l_2 & l_3 \\ 
l^{\prime} & l^{\prime\prime} & l
\end{array}
\right\} \left( -1\right) ^{l+l^{\prime }+l^{\prime \prime }}
\end{eqnarray}
If $l^{\prime \prime }=0,$ then the only non-zero term is 
\begin{equation}
Q_{l_1 l_3 0}^{l_1l_2l_3} =\sqrt{\frac{\left( 2l_2+1\right) (4\pi )^3}{%
(2l_1+1)(2l_3+1)}}\left( 
\begin{array}{ccc}
l_1 & l_2 & l_3 \\ 
0 & 0 & 0
\end{array}
\right)
\end{equation}
If $l^{\prime \prime }=1,$ $Q_{ll^{\prime }1}^{l_1l_2l_3}$ is zero unless $%
l=l_1\pm 1$ and $l^{\prime }=l_3\pm 1.$ Thus, there are only four terms in
the sum in equation (27).

We can evaluate these terms by noting that 
\begin{equation}
\left( 
\begin{array}{ccc}
l_1-1 & l_2 & l_3-1 \\ 
0 & 0 & 0
\end{array}
\right) =\sqrt{\frac{(l_1+l_2+l_3+1)(l_1-l_2+l_3)}{%
(l_1+l_2+l_3)(l_1-l_2+l_3-1)}}\left( 
\begin{array}{ccc}
l_1 & l_2 & l_3 \\ 
0 & 0 & 0
\end{array}
\right)
\end{equation}

\begin{equation}
\left( 
\begin{array}{ccc}
l_1 & l_1-1 & 1 \\ 
0 & 0 & 0
\end{array}
\right) =-\sqrt{\frac{l_1}{(2l_1+1)(2l_1-1)}}
\end{equation}
and 
\begin{equation}
\left\{ 
\begin{array}{ccc}
l_1 & l_2 & l_3 \\ 
l_3-1 & 1 & l_1-1
\end{array}
\right\} =\sqrt{\frac{%
(l_1+l_2+l_3+1)(l_1+l_2+l_3)(l_1-l_2+l_3-1)(l_1-l_2+l_3)}{%
(2l_1+1)2l_1(2l_1-1)(2l_3+1)2l_3(2l_3-1)}}
\end{equation}
Thus we can reduce the product of Wigner symbols to a simpler form: 
\begin{equation}
Q_{l_1-1, l_3-1, 1}^{l_1l_2l_3}=\frac{(l_1+l_2+l_3+1)(l_1-l_2+l_3)}{%
2(2l_1-1)(2l_3-1)}Q_{l_1 l_3 0}^{l_1 l_2 l_3}  \label{Qmm}
\end{equation}
Similarly, 
\begin{equation}
Q_{l_1+1, l_3-1,1}^{l_1l_2l_3}=\frac{(l_1+l_2-l_3+1)(-l_1+l_2+l_3)}{%
2(2l_1+3)(2l_3-1)}Q_{l_1 l_3 0}^{l_1l_2l_3},  \label{Qpm}
\end{equation}
\begin{equation}
Q_{l_1-1, l_3+1,1}^{l_1l_2l_3}=\frac{(-l_1+l_2+l_3+1)(l_1+l_2-l_3)}{%
2(2l_1-1)(2l_3+3)}Q_{l_1 l_3 0}^{l_1l_2l_3},  \label{Qmp}
\end{equation}
and 
\begin{equation}
Q_{l_1+1, l_3+1, 1}^{l_1l_2l_3}=\frac{(l_1+l_2+l_3+2)(l_1-l_2+l_3+1)}{%
2(2l_1+3)(2l_3+3)}Q_{l_1 l_3 0}^{l_1l_2l_3},  \label{Qpp}
\end{equation}
Note that 
\begin{equation}
\sum_{ll^{\prime }}Q_{ll^{\prime }1}^{l_1l_2l_3}(2l+1)(2l^{\prime
}+1)=(2l_1+1)(2l_3+1)Q_{l_1 l_3 0}^{l_1l_2l_3},  \label{Wigner}
\end{equation}

\end{document}